\documentclass{article}
\usepackage{url}

\usepackage{amsmath}
\usepackage{times}
\usepackage{bm}
\usepackage{natbib}
\usepackage{color}
\usepackage{multirow}
\usepackage{graphicx}
\usepackage{varwidth}
\usepackage{float}
\usepackage[table,x11names]{xcolor}
\definecolor{light-gray}{gray}{0.95}
\usepackage{amssymb}
\usepackage{multirow}
\usepackage{geometry}
\usepackage{setspace}
 \newcommand{\prob}[1]{\ensuremath{\mbox{\textup{pr}}\left(#1\right)}}

\newcommand{\by}{\ensuremath{y}}

\usepackage{amsmath} 
\usepackage{amsthm}              
      \newtheorem{assumption}{Assumption}
      
      \newtheorem{theorem}{Theorem}

\usepackage[plain,noend]{algorithm2e}

\begin{document}

\title{A robustified posterior for Bayesian inference on a large number of parallel effects}


\author{J. G. LIAO$^\ast$, ARTHUR BERG, TIMOTHY L. MCMURRY\\  \\ \textit{Division of Biostatistics \& Bioinformatics}\\ \textit{Pennsylvania State University}\\ \\ \textit{Hershey, PA 17033,  USA}\\ \\  jliao@phs.psu.edu}

\date{}

\maketitle

\begin{abstract}
{Many modern experiments, such as microarray gene expression and genome-wide association studies, present the problem of estimating a large number of parallel effects.  Bayesian inference is a popular approach for analyzing such data by modeling the large number of unknown parameters as random effects from a common prior distribution. However, misspecification of the prior distribution can lead to erroneous estimates of the random effects, especially for the largest and most interesting effects.  This paper has two aims. First, we propose a robustified posterior distribution for a parametric Bayesian hierarchical model that can substantially reduce the impact of a misspecified prior. Second, we conduct a systematic comparison of the standard parametric posterior, the proposed robustified parametric posterior, and a nonparametric Bayesian posterior which uses a Dirichlet process mixture prior. The proposed robustifed posterior when combined with a flexible parametric prior can be a superior alternative to nonparametric Bayesian methods.}
\end{abstract}

\section{Introduction}

In the past decades, new technologies such as gene microarray and genome-wide association studies have fundamentally changed the landscape of biomedical research. Instead of studying one gene or one single nucleotide polymorphism (SNP) at a time, these technologies allow us to study thousands of genes or SNPs simultaneously. Bayesian approaches have proven effective for analyzing such data by modeling a large number of parallel parameters for individual genes or SNPs as random effects from a common prior. Bayesian methods can improve inference by borrowing information from other genes and by incorporating useful structure such as modeling a large proportion of the genes or SNPs as having no effect on the outcome.  This approach automatically adjusts for multiple comparisons and selection bias inherent in the large-scale data setting (Johnstone and Silverman, 2004; Efron, 2010). 

A canonical model for this data structure is
\begin{equation}
\label{eq:canonical}
y_i=\theta_i+\varepsilon_i,\quad i=1,\ldots,p,
\end{equation}
where $y_i$ is the observed measurement, $\theta_i$ is the unknown true parameter of interest for the $i^\text{th}$ gene or SNP, and $\varepsilon_i$ is an unobserved random error.   For many applications, there is often little information to distinguish one $\theta_i$ from the others before the data is collected, and in such situations, we can consider the $\theta_i$ exchangeable \citep[Chapter 5]{Gelman:2014aa}. For this paper, we shall treat $\theta_i$ as random effects drawn from a density $\pi_0$, and consider $\pi_0$ as a smooth or limiting form of the empirical distribution of the underlying $\theta_1,\ldots,\theta_p$ to be estimated. Letting $\theta=(\theta_1,\ldots,\theta_p)$ and $y=(y_1,\ldots,y_p)$, a suitable posterior distribution that facilitates the inference about $\theta$ is 
\[
f(\theta\mid y,\pi_0)\propto f(y\mid\theta)\pi_0(\theta).
\]
We would like to approximate $f(\theta\mid y,\pi_0)$ as closely as possible in our Bayesian inference.  

In practice, however, $\pi_0$ is usually unknown. A standard way to take advantage of such data structure is through a parametric Bayesian hierarchical model \citep[Chapter 5]{Gelman:2014aa} as follows:
\begin{equation}
\label{eq:heir}
\begin{split}
y_i  \mid \theta_i&\sim f(y_i\mid\theta_i)\\
\theta_i \mid \eta &\sim f(\theta_i \mid \eta)\\
\eta&\sim f(\eta),
\end{split}
\end{equation}
 where $f(y_i|\theta_i)$ is defined by the desnsity of error $\varepsilon_i$ in \eqref{eq:canonical},  $y_1,...,y_p$ are independent given $\theta_1,\ldots,\theta_p$, and $\theta_1,\ldots,\theta_p$ are independent given $\eta$. This approach can easily incorporate prior information about the structure of $\pi_0$ for improved inference about $\theta_i$. For example, we can choose the working prior  $f(\theta_i|\eta)$ to be a Laplace family with scale parameter $\eta$ if we believe that $\pi_0$ is unimodal and long-tailed  and a normal family if short-tailed.  When the shape of $f(\theta_i|\eta)$ is misspecified and severely deviated from $\pi_0$, however, it can lead to inferior inference. For example, excessive shrinkage and therefore bias can occur if a working prior $f(\theta_i\mid\eta)$ has much shorter tails than $\pi_0$. This misspecification of working prior can be a serious concern because, while some information about $\pi_0$ may be available in a particular application, the information is typically insufficient to determine how heavy the tails should be, which can substantially influence the extremal effects of $\theta_i$, usually the effects of most practical interest. 

Alternatively, nonparametric and semi-parametric Bayesian methods   \citep{Muralidharan:2010aa, Martin:2012aa, Bogdan:2008aa, Muller:2013aa, Do:2005aa, Kim:2009aa} can be used that impose minimum structure on $\pi_0$. A popular approach is to specify the working prior for $\theta_i$ as generated from a Dirichlet process mixture, as described in more detail in Section 4. Although such a working prior generally performs reasonably well for a wide range of $\pi_0$, it may not be optimal due to its weaker prior information. Additionally, its focus on flexibility of the model can make it difficult for a statistician to incorporate useful prior information \citep{Hoff:2013ab,OHagan:2013aa,Carlin:2013aa}. However, to our knowledge, no systematic comparison of parametric and nonparametric approaches, via simulation study or real data-sets, is available in the setting of a large number of effects ($p$ in the order of thousands) to give empirical guidance in real applications.

This paper has two aims. First, we propose a simple new method to robustify the posterior distribution of $\theta_i$ for parametric hierarchical model \eqref{eq:heir} by utilizing the asymptotic behavior of order statistics. A unique feature of the proposed method is that it can substantially improve the posterior distribution when $f(\theta_i|\eta)$ is misspecified without affecting the posterior under a correctly specified working prior. Therefore, it can be broadly used.  Second, we conduct a systematic performance comparison of the standard parametric posterior, the proposed robustified parametric posterior, and a nonparametric Bayesian posterior which uses a Dirichlet process mixture prior with a normal base distribution. Our study shows that while the nonparametric Bayesian method does provide reasonable performance under different forms of $\pi_0$, it can perform poorly when $\pi_0$ is severely deviated from normal, possibly due to its normal base distribution in the Dirichlet process mixture prior. The proposed robustifed posterior when combined with a flexible parametric prior can be a superior alternative to nonparametric Bayesian methods.

\section{The Robustified Posterior}   \label{sec.post}

We now present the key result of how to robustify the standard posterior distribution for a general working prior density $\pi$ given by
\begin{equation}
\label{eq:r}
f(\theta\mid y,\pi)\propto f(y\mid\theta)\pi(\theta).
\end{equation}
  In order to develop the robustified version denoted by $f_{\text{robust}}(\theta\mid y,\pi)$, we need  the following assumption, which is reasonable if $y_i$ is to be a measurement of $\theta_i$.

\begin{assumption}
\label{as:mono}
We assume the distribution of $y_i \mid \theta_i$ is strictly stochastically increasing in $\theta_i$. 
\end{assumption}
Let $\Phi_i$ be the cumulative distribution function of the errors $\varepsilon_i=y_i-\theta_i$ in \eqref{eq:canonical} and let $\phi_i$ be the corresponding  density. We then have $f(y_i\mid \theta_i)=\phi(y_i-\theta_i \mid \theta_i)$. Let  
\begin{equation}
\label{eq:quantile}
u_i=\Phi_i(y_i-\theta_i \mid \theta_i)
\end{equation}
be  the quantiles of error $\varepsilon_i$. Here we allow the distribution of $\varepsilon_i$ 
to depend on $\theta_i$ in order to be more general. It follows immediately that $u_i\stackrel{\text{iid}}{\sim}\text{unif}(0,1)$.
 For a given $y_i$, $\Phi_i(y_i-\theta_i \mid \theta_i)$ is a strictly decreasing function of $\theta_i$ by Assumption 1 and therefore $\theta_i$ can be written as a function of $u_i$: $\theta_i=g_i(u_i)$. We will reformulate the posterior of $\theta_i$, given in \eqref{eq:r}, in terms of the quantiles of error $u_i$. Then the change-of-variables formula rewrites the posterior \eqref{eq:r} as
\begin{equation} \label{eq.udens} 
f(u \mid y, \pi) \propto \pi\left(g(u)\right) \phi(y-g(u) \mid g(u)) \left|\frac{d \theta}{d u}\right|.
\end{equation} 
where $u=(u_1,\ldots,u_p)$, $g(u)=(g_1(u_1),\ldots,g_p(u_p))$,  and $\phi=(\phi_1,\ldots,\phi_p)$.

An interesting special case occurs when the distribution of
$\varepsilon_i$ does not depend on $\theta_i$.  In this case, $\theta_i=g_i(u_i)=y_i-\Phi_i^{-1}(u_i)$ and $
\left|\frac{d u}{d \theta}\right| =
\prod_{i=1}^p\phi_i(y_i-\theta_i)$.  Therefore the posterior distribution~\eqref{eq.udens}
has the particularly simple form 
\begin{equation}
\label{eq:special}
f(u \mid y, \pi) = \pi\left(y -
\Phi^{-1} (u) \right). 
\end{equation}

We now return to the general case of \eqref{eq.udens}.  Let $\tilde u=(u_{[1]}, \ldots,
u_{[p]})$ denote the order statistics of $u=(u_1, \ldots, u_p)$.  Then the posterior distribution of $u$ under the working prior $\pi$ has the decomposition
\begin{equation} \label{eq.postdecomp} f(u \mid \by, \pi) =
f(u \mid \tilde u, \by, \pi) f(\tilde u \mid \by, \pi). \end{equation} If the working prior $\pi$ is
misspecified, both factors in decomposition~\eqref{eq.postdecomp} can be distorted from their corresponding distribution under the correctly specified prior $\pi_0$.  The key idea of our proposed robustified posterior distribution is to replace $f(\tilde u \mid \by, \pi)$ in  \eqref{eq.postdecomp} with its the asymptotic limit $f(\tilde u \mid \by, \pi_0)$, which turns out not to depend on $\pi_0$. In the rest of this section, we shall assume that $y$ is generated under model \eqref{eq:canonical} with $\theta_i\sim\pi_0$. In what follows, we show the asymptotic limit of $f(\tilde u \mid \by,\pi_0)$ is available without knowledge of the correct prior $\pi_0$.  More specifically, $f(\tilde u \mid \by,\pi_0)$ converges to the discrete uniform distribution on $\{\frac{1}{p+1},\ldots,\frac{p}{p+1}\}$.  The key insight is that when $p$ is large
and under the correct $\pi_0$, $\tilde u$ is well
approximated by the quantiles of the uniform distribution on $[0,1]$;
this is the same rationale as justifies the widely used QQ-plot for distribution checking.  We formalize this in the following theorem (proof provided in the Appendix).
\begin{theorem}
\label{thm:sup}
Let  $y$ be generated under model \eqref{eq:canonical} with $\theta_i\sim\pi_0$.   
Let $\tilde u=(u_{[1]}, \ldots,
u_{[p]})$ denote the order statistics of $u=(u_1, \ldots, u_p)$ drawn from  $f(\tilde u\mid y,\pi_0)$. Then
\[
\sup\left|u_{[i]}-\frac{i}{p+1}\right|\stackrel{p}{\longrightarrow}0,
\]
 except on a small subset of $y$ whose probability can be made as small as any   $\delta>0$. 
\end{theorem}

We therefore propose to fix the $\tilde u$ in $f(u \mid \tilde u, y, \pi)$ in the right hand side in \eqref{eq.postdecomp} to be its  asymptotic limit $u_{[i]}=i/(p+1)$ derived under $f(\tilde u \mid y,\pi_0)$ and define our robustified posterior as 
\begin{equation} 
\label{eq.robpost} 
f_{\mbox{robust}}(u \mid \by, \pi) =
f\left(u \mid u_{[1]} = \frac{1}{p+1}, \ldots, u_{[p]} =
\frac{p}{p+1}, \by, \pi \right). 
\end{equation} 
In the robust
posterior \eqref{eq.robpost}, the sample space of $u$, to be denoted by $\varGamma$, consists of the $p!$ permutations of $u_{[1]},\ldots,u_{[p]}$, where $u_{[1]} = \frac{1}{p+1}, \ldots, u_{[p]} =
\frac{p}{p+1}$.   
This gives the following explicit form of \eqref{eq.robpost}
\begin{equation} 
\label{eq:robust}
f_{\mbox{robust}}(u \mid \by, \pi) = c(y) f(u \mid \by, \pi)
\end{equation} 
on $u\in \varGamma$, where $c^{-1}(y)=\sum_{u\in \varGamma}f(u \mid \by, \pi)$.  Note that this approach of robustifying the standard posterior through truncation to the discrete space $\varGamma$ can be applied to the posterior for any general Bayesian model including hierarchical Bayes and empirical Bayes. Using relationship $\theta=g(u)$ as defined at the beginning of this section, we can easily map $f_{\mbox{robust}}(u \mid \by, \pi)$ back to $\theta$ as $f_{\mbox{robust}}(\theta \mid \by, \pi)$. In particular, $\theta=g(u)\sim f_{\mbox{robust}}(\theta \mid \by, \pi)$ if $u\sim f_{\mbox{robust}}(u \mid \by, \pi)$.

Based on the discussion above,  the robustified posterior \eqref{eq:robust} is effective over the standard posterior when the misspecified $\pi$ causes   $f(\tilde u\mid\by, \pi)$ to deviate from $f(\tilde u\mid\by,\pi_0)$.   This can happen when the working prior over-shrinks due to underspecification of the working prior $\pi$ or shrinks in the wrong direction due to a location shift.  On the other hand, it does not improve inference if the misspecified $\pi$ primarily affects $f(u \mid \tilde u, \by, \pi)$ in \eqref{eq.postdecomp}, which happens when the working prior is too diffuse and therefore under-shrinks.  

For hierarchical model \eqref{eq:heir}, let 
\begin{equation}
\label{eq:pi_h}
\pi_h(\theta)=\int f(\theta\mid\eta)f(\eta)\,d\eta
\end{equation}
be the prior of $\theta$ after integrating out $\eta$. We can use Theorem \ref{thm:sup} to improve the inference of  hierarchical model \eqref{eq:heir} by replacing $f(\theta\mid y, \pi_h)$, as defined in \eqref{eq:r}, by $f_{\text{robust}}(\theta\mid y,\pi_h)$, which is, however, computationally prohibitive because the mixture density $\pi_h$ given in \eqref{eq:pi_h} is very expensive to evaluate. To get around this this computational limitations, note that the standard posterior $f(\theta\mid y, \pi_h)$ can be simulated as the stationary distribution of Gibbs sampler 
\[
\begin{cases}
\theta^{[j]}\sim f(\theta\mid \eta^{[j-1]}, y)\\
\eta^{[j]}\sim f(\eta\mid \theta^{[j]}, y).
\end{cases}
\]
Now a robustified version of the component $f(\theta\mid \eta, y)$, $f_{\text{robust}}(\theta\mid \eta, y)$, can be constructed as $f_{\text{robust}}(\theta\mid y,\pi)$,  where $\pi(\theta)=f(\theta\mid\eta)$, which is much less computationally intensive because $f(\theta\mid\eta)$ is usually a simple parametric distribution. We use it to  construct a robustified Gibbs sampler
\begin{equation}
\label{eq:RobustGibbs}
\begin{cases}
\theta^{[j]}\sim f_{\text{robust}}(\theta\mid \eta^{[j-1]}, y),\\
\eta^{[j]}\sim f(\eta\mid  \theta^{[j]}, y),
\end{cases}
\end{equation}
which can be simulated from rapidly. We shall show in Section 3 that this new Gibbs sampler has a stationary distribution. Our proposed robustified posterior is defined as the the stationary distribution of $\theta$ for this  robustified Gibbs sampler, which can replace the  standard posterior $f(\theta\mid y, \pi_h)$ for improved Bayesian inference.


\section{A Markov Chain Monte Carlo Algorithm} \label{sec.mcmc}

We now describe a random walk Metropolis-Hasting algorithm \citep[e.g.][Chapter 6]{Robert:2009aa} to sample from $f_{\mbox{robust}}(u\mid y,\pi)$ in \eqref{eq:robust}, whose sample space $\varGamma$ consists of $p!$ permutations of $\{\frac{1}{p+1},\ldots,\frac{p}{p+1}\}$.  Let $u \in \varGamma$ be the current position of the Markov chain.  Randomly select $k$ locations in $u$ and then randomly permute these $k$ elements at these locations. Let the resulting $u$ be the candidate $u^c$.  The parameter $k$ plays the role of a step size in a continuous random walk.  It is easy to see that this algorithm is symmetric; the probability of generating $u^c$ from $u$ is the same as the probability of generate $u$ from $u^c$.  Therefore, the Metropolis-Hastings algorithm accepts $u^c$ with probability
\[ \alpha \equiv \min \left(1, 
\frac{f(u^c \mid y, \pi)}{f(u \mid y, \pi)} \right). \]

Due to the enormous number of points in the discrete sample space, $\varGamma$, an effective Metropolis-Hasting algorithm must start the chain from a well supported point and must be able to control the distance of the proposal $u^c$ from current position $u$. To address this, we propose the following enhancement. Let $q_i=\Phi_i(y_i\mid \theta_i=0)$ be the $p$-value of testing $H_0: \theta=0$ versus $H_1: \theta_i<0$, which can serve as an approximation of the unknown $u_i=\Phi_i(y_i\mid \theta_i)$. Note also that $1-q_i$ is the $p$-value for testing $H_0: \theta_i=0$ versus $H_1: \theta_i>0$.  Now reorder $y_1,\ldots, y_p$ by the value of $q_i$. To simplify notation, the reordered sequence will still be denoted as $y_1,\ldots, y_p$ but now satisfies 
\[
q_1\le q_2\le\cdots\le q_p
\]
Since $q_i$ is an approximation of $u_i$, the underlying $u_1, u_2,\ldots,u_p$ for the reordered $y_1,\ldots,y_p$ should approximately follow the same increasing pattern as $i$ increases. We propose to apply the above random walk Metropolis-Hasting algorithm to the reordered dataset $y_1,\ldots,y_n$ with two benefits. first, we start the chain at a well-supported point $u=(1/(p+1), \ldots, p/(p+1))$. Second, in generating proposal $u^c$ from $u$, we randomly select $k$ consecutive positions in $u$ and then randomly permute elements at these locations. Because the consecutive elements in $u_i$ now generally have similar values, we can easily control the probability of accepting $u^c$ to be around 25\% as recommended in \citet[Section 6.6]{Robert:2009aa} by selecting an appropriate $k$. This enhancement drastically improves the sampling efficiency in our experience.

We can now easily implement robustified Gibbs sampler  \eqref{eq:RobustGibbs}. Drawing from $f(\eta\mid \theta^{[j]})$ is usually straightforward. The above Metropolis-Hasting algorithm can be used to sample from $f_{\text{robust}}(u\mid \eta^{[j-1]},y)$ and therefore $f_{\text{robust}}(\theta\mid \eta^{[j-1]},y)$. Note that $u$ has a finite sample space of $\Gamma$ and the chain is irreducible so long as $f(u\mid \eta,y)>0$ for all $u\in \Gamma$ under any given $\eta$. It follows that this robustified Gibbs sampler has a unique stationary distribution, whose density can be written down explicitly from the two conditional distributions in the Gibbs sampler by the Hammersley-Clifford theorem \citep{Besag:1974aa}.

\section{Comparison of Performance} \label{sec.sim}

A comprehensive simulation study is perfromed to compare the performance of three sets of procedures: standard parametric posterior, the robustified parametric posterior, a nonparametric Bayesian using Dirichlet process mixture prior. The simulation study is conducted as follows.

\begin{description}
 \item[Step 1:] For $p=1000$ and $p=2000$, generate $y_i = \theta_i +
\varepsilon_i$ (for $i=1,\ldots,p$), where $\theta_i \stackrel{\text{iid}}{\sim} \pi_0$ and
$\varepsilon_i \stackrel{\text{iid}}{\sim} N(0,1)$. 

\item[Step 2:] 
Reorder data $y_1,\ldots,y_p$ by the values of $q_1,\ldots,q_p$ as described in Section 3.  Arrange the generated $\theta_1,\ldots,\theta_p$ in the corresponding order.

\item[Step 3:] For dataset $y_1,\ldots,y_p$, compute the posterior means $\hat\theta_i$ and the estimation errors $\hat\theta_i-\theta_i$ ($i=1,\ldots,p$) using seven estimation methods described below.

\end{description}

Steps 1 through 3 are repeated 100 times for the results reported here.  In this simulation study, three forms of $\pi_0$ are used.  The first form, $\pi_0^N$, is a normal distribution $N(0, 2^2)$, which serves as an example of a light-tailed distribution.  The second form, $\pi_0^t$, is the scaled $t$-distribution with five degrees of freedom and a standard deviation of two, which represents a heavy-tailed distribution.  The third form, $\pi_0^h$, is given by 
\[
\pi_0^h=0.9\pi_0^{N,\text{trunc}}+0.1\pi_0^{t,\text{trunc}},
\] 
where $\pi_0^{N,\text{trunc}}$ is $\pi_0^N$  truncated to interval $[-4,4]$ and $\pi_{0}^{t,\text{trunc}}$ is $\pi_0^t$ truncated to $(-\infty,-4)\cup(4,\infty)$.  This hybrid distribution has the form of $\pi_0^N$ in the middle and the form of $\pi_0^t$ in the tails. 


We now describe the seven estimation methods used in this simulation. 
\begin{description}
\item[Method 1 (Laplace).]  Standard  posterior for hierarchical model \eqref{eq:heir} with Laplace working prior:   $\varepsilon_i \stackrel{\text{iid}}{\sim} N(0,1)$, $\theta_i\mid \eta_1\sim\mathrm{Laplace}(0,\eta_1)$ with scale parameter $\eta_1\sim \text{Unif}(0,35.35)$.   
\item[Method 2 (R Laplace).] Robustified posterior of Method 1.
\item[Method 3 (Normal).]
Standard  posterior for hierarchical model \eqref{eq:heir} with normal working prior:  $\theta_i\mid \eta_2\sim \mathcal{N}(0,\eta_{2}^2)$ with $\eta_2\sim \text{Unif}(0,50)$. Note the variances of $\mathcal{N}(0,\eta_{1}^2)$ with $\eta_1=35.35$ and $\mathrm{Laplace}(0,\eta_2)$ with $\eta_2=50$ are the same.
\item[Method 4 (R Normal).] Robustified posterior of Method 3.  
\item[Method 5 (Mixture).] Standard posterior for hierarchical model \eqref{eq:heir} with a mixture working prior: 
\[
\theta_1,\ldots,\theta_p\mid \lambda,\eta_1,\eta_2\sim \lambda\prod_{i=1}^p \mathrm{Laplace}(\theta_i\mid 0, \eta_1)+(1-\lambda)\prod_{i=1}^p\mathcal{N}(\theta_i\mid 0,\eta_2^2)
\]
with $\lambda\sim\text{Unif}(0,1)$, and $\eta_1$ and $\eta_2$ distributed as in Methods 1 and 3. 
\item[Method 6 (R Mixture).] Robustified posterior of Method 5.
\item[Method 7 (DP).] Nonparametric Bayes with Dirichlet process mixture prior:
\[
\begin{split}
&\theta_i\mid G,\sigma\sim \int \mathcal{N}(\mu,\sigma^2)\,d G(\mu)\\
&\sigma^2\sim \textrm{Inv-Gamma}(3,5)\\
&G\mid G_0,\alpha\sim DP(\alpha,G_0), \text{where } \alpha=1, G_0=N(0,\sigma_b^2)\\
&\sigma_b^2\sim \textrm{Inv-Gamma}(5,20).\\
\end{split}
\]
In this formulation, $\alpha$ is the scaling parameter and $G_0$ is the base distribution of the Dirichlet process. The shape and scale parameters in the two inverse Gamma distributions are chosen so that $\sigma^2$ and $\sigma_b^2$ have sufficient variability for a flexible model.   
\end{description}

For $p=1000$, figures 1 presents the distribution of $\hat\theta_1-\theta_1$ and $\theta_{p}-\hat\theta_{p}$ over 100 replications under each of $\pi_0^N$, $\pi_0^t$, $\pi_0^h$, and for each of the 7 estimation methods using boxplots. Figure 2 presents the same results for $p=2000$.  Note that we can combine $\hat\theta_i-\theta_i$ and $\theta_{p+1-i}-\hat\theta_{p+1-i}$ into one boxplot to save space because they have the same distribution due to the symmetries of $\pi_0$ and the working priors. Concentration of the distribution around 0 in a boxplot  represents a good estimation method while concentration below and above 0  represent under-shrinkage and over-shrinkage respectively. Note also that $\theta_1,...,\theta_p$ have been re-ordered in Step 2 above by the $q_1,\ldots,q_p$ values. Therefore,  $\theta_1$ and $\theta_{1000}$  are the two most extremal effects usually of the greatest practical interest.  

We now summarize the performance of the 7 estimation methods as reflected in figures 1 and 2. Method 1  under-shrinks considerably under $\pi_0^{N}$ as expected because the working Laplace prior has much heavier tails. It works well under heavier-tailed  $\pi_0^{t}$ and $\pi_0^{h}$. Method 2 offers small improvement over method 1 for $p=2000$. Method 3 works very well under $\pi_0^N$ but severely over-shrinks under $\pi_0^t$ and $\pi_0^h$. Method 4 substantially improves over Method 3 under $\pi_0^t$ and $\pi_0^h$ but still performs poorly  due to the restrictive short tails in the working prior. Method 5, with its flexible mixture working prior, has low bias across all three forms of $\pi_0$ but suffers from increased variation.  Method 6 offers considerable improvement over Method 5 under $\pi_0^t$ and $\pi_0^h$, in particular by reducing the extreme estimation errors.  The nonparametric Method 7 performs  well under $\pi_0^N$, less well under $\pi_0^h$, and poorly under $\pi_0^t$, possibly due to its normal base-distribution.  Changing the scaling parameter $\alpha$ to 1/2 to increase the variability of the Dirichlet process does not improve its performance (data not shown). Nevertheless, Method 7 never performs very badly across all three forms of $\pi_0$, demonstrating the relative robustness of the nonparametric method. Overall, Method 6, which combines the robustified posterior with a reasonably flexible parametric working prior, has the best performance. It is also shown that the improvement provided by robustification becomes more dramatic when $p$ is increased from 1000 to 2000.


There are a few cases in figures 1 and 2 in which the robustified posterior provides little improvement over the standard posterior. This can happen when the working prior is already optimal or close to optimal, such as Method 1 under $\pi_0^t$ and Method 3 under $\pi_0^N$. It can also happen when the working prior has longer tails than $\pi_0$ as noted in Section 2, such as Method 1 under $\pi_0^N$. 

Finally, Tables 1 and 2 give, for $p=1000$ and $p=2000$ respectively, the mean square error of each estimation method as the average of $(\hat\theta_i-\theta_i)^2$ and $(\theta_{p+1-i}-\hat\theta_{p+1-i})^2$ from the 100 replications for $i=1, 2, 3$. The performance ranking of the 7 methods summarized above for $i=1$ in figures 1 and 2 is  still generally valid for $i=2, 3$ but the difference between different methods is much smaller.    

The standard posterior estimates in Methods 1, 3 and 5 
 are computed using RStan \citep{Carpenter:2016aa}. The  robustified posterior estimates for Methods 2, 4, and 6 are obtained by our own R code. Function \texttt{DPMmeta} in the DPpackage \citep{Jara:2011aa} is used for the nonparametric Bayesian estimation in Method 7. Our complete R code for this simulation study is available at \url{http://sites.google.com/site/jiangangliao}.

\begin{figure}[H] 
\label{fig1}
\begin{center} 
\includegraphics[width=.7\textwidth]{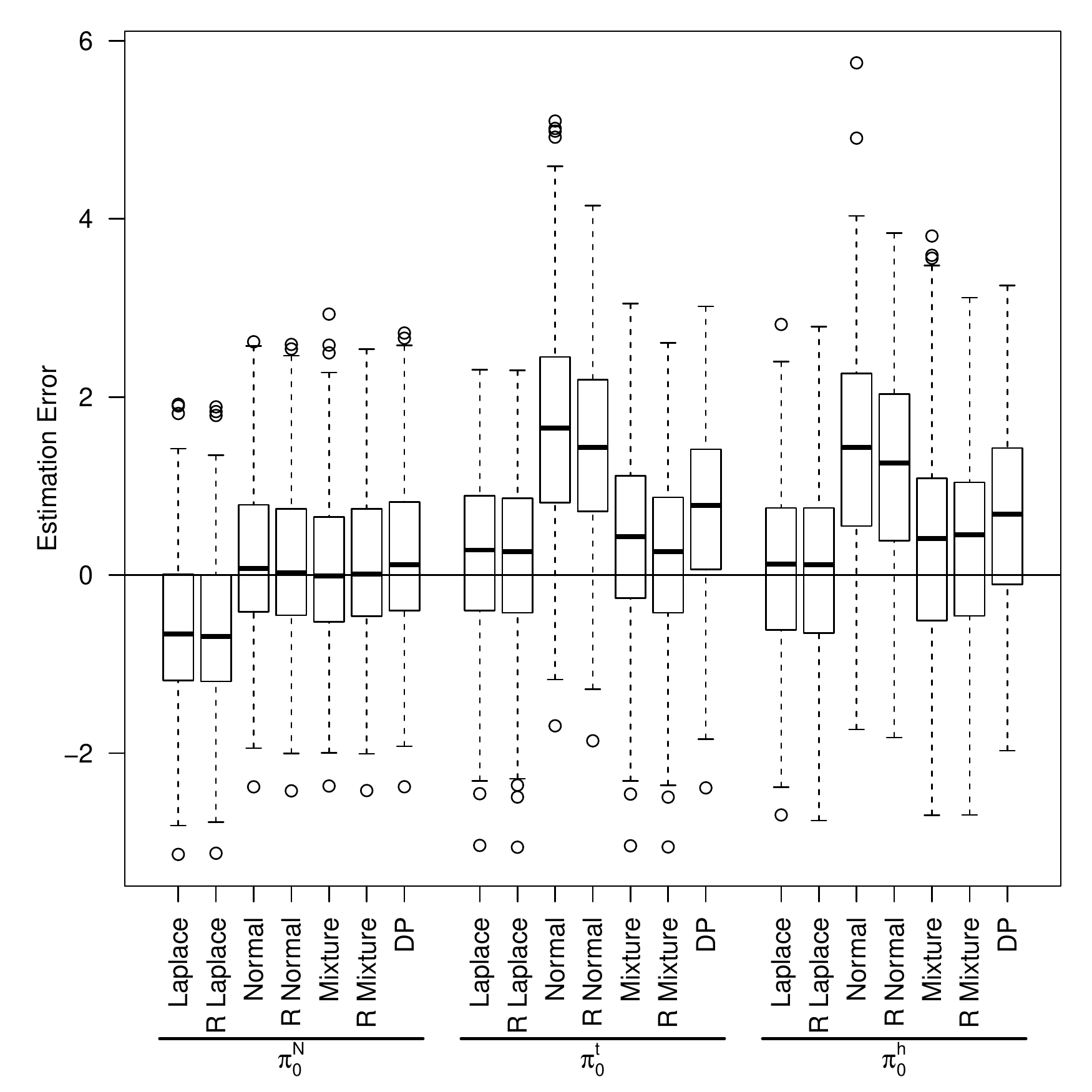} 
\caption{$p$=1000. Boxplots of the estimation error of the most extreme random effects; $\theta_1-\hat\theta_1$ and $\theta_{1000}-\hat\theta_{1000}$. } 
\end{center} 
\end{figure}

\begin{table}[H]
\label{tab1}
\caption{$p=1000$. Mean square error of $\hat\theta_i$ and $\hat\theta_{1001-i}$ for the seven methods under $\pi_0^N$, $\pi_0^t$, and $\pi_0^h$. } 
\centering
\begin{tabular}{|c|l|rr|rr|rr|r|}
  \hline
$i$ & $\pi_0$ & Laplace & R Laplace & Normal & R Normal & Mixture & R Mixture & DP \\ 
  \hline
 & $\pi_0^N$ & 1.25 & 1.29 & 0.85 & 0.84 & 0.90 & 0.84 & 0.88 \\ 
  1 & $\pi_0^t$ & 1.02 & 1.02 & 4.29 & 3.25 & 1.31 & 1.03 & 1.61 \\ 
   & $\pi_0^h$ & 1.09 & 1.08 & 3.57 & 2.81 & 1.59 & 1.36 & 1.53 \\ 
   \hline
 & $\pi_0^N$ & 1.31 & 1.34 & 0.79 & 0.79 & 0.89 & 0.80 & 0.80 \\ 
  2 & $\pi_0^t$ & 1.13 & 1.12 & 2.30 & 2.05 & 1.25 & 1.14 & 1.23 \\ 
   & $\pi_0^h$ & 1.26 & 1.25 & 2.28 & 2.09 & 1.46 & 1.39 & 1.41 \\ 
   \hline
 & $\pi_0^N$ & 1.12 & 1.14 & 0.85 & 0.84 & 0.88 & 0.83 & 0.84 \\ 
  3 & $\pi_0^t$ & 1.09 & 1.09 & 1.90 & 1.75 & 1.14 & 1.10 & 1.17 \\ 
   & $\pi_0^h$ & 1.08 & 1.08 & 1.65 & 1.56 & 1.14 & 1.10 & 1.12 \\ 
\hline  
\end{tabular}
\end{table}

\begin{figure}[H] 
\label{fig2}
\begin{center} 
\includegraphics[width=.7\textwidth]{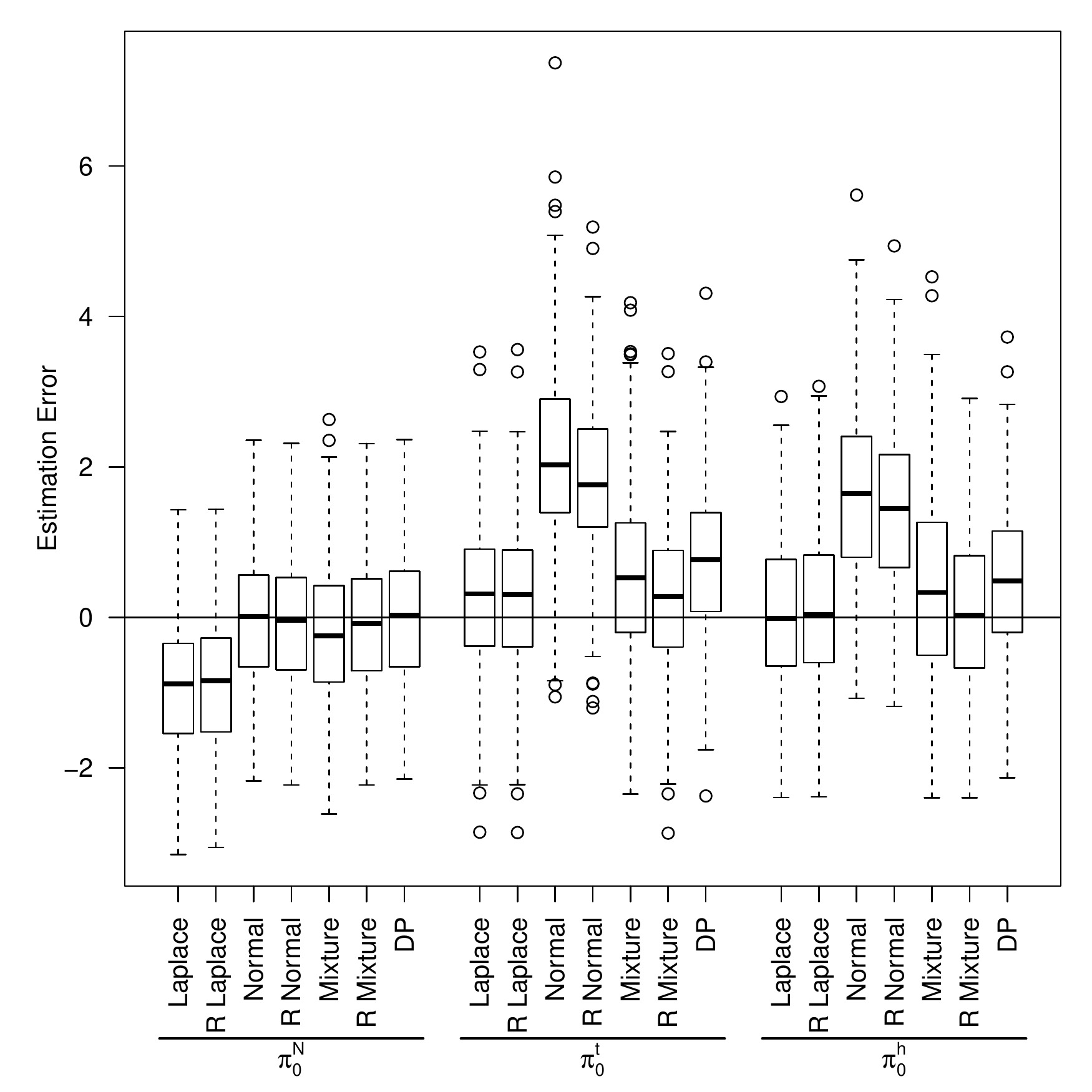} 
\caption{$p$=2000.  Boxplots of the estimation error of the most extreme random effects; $\theta_1-\hat\theta_1$ and $\theta_{2000}-\hat\theta_{2000}$. } 
\end{center} 
\end{figure}

\begin{table}[H]
\label{tab2}
\caption{$p$=2000.  Mean square error of $\hat\theta_i$ and $\hat\theta_{2001-i}$ for the seven methods under $\pi_0^N$, $\pi_0^t$, and $\pi_0^h$. } 
\centering
\begin{tabular}{|c|l|rr|rr|rr|r|}
  \hline
$i$ & $\pi_0$ & Laplace & R Laplace & Normal & R Normal & Mixture & R Mixture & DP \\ 
  \hline
 & $\pi_0^N$ & 1.56 & 1.50 & 0.74 & 0.74 & 0.94 & 0.76 & 0.75 \\   
  1 & $\pi_0^t$ & 1.07 & 1.07 & 6.08 & 4.33 & 1.79 & 1.06 & 1.60 \\ 
   & $\pi_0^h$ & 0.97 & 0.97 & 4.20 & 3.21 & 1.81 & 0.98 & 1.34 \\ 
   \hline
 & $\pi_0^N$ & 1.38 & 1.33 & 0.85 & 0.85 & 1.02 & 0.87 & 0.87 \\ 
  2 & $\pi_0^t$ & 1.33 & 1.33 & 3.51 & 3.02 & 1.87 & 1.33 & 1.48 \\ 
   & $\pi_0^h$ & 1.18 & 1.19 & 2.82 & 2.56 & 1.52 & 1.23 & 1.32 \\ 
   \hline
 & $\pi_0^N$ & 1.40 & 1.35 & 0.99 & 0.99 & 1.15 & 1.01 & 1.00 \\ 
  3 & $\pi_0^t$ & 1.15 & 1.16 & 2.67 & 2.41 & 1.30 & 1.15 & 1.18 \\ 
   & $\pi_0^h$ & 1.02 & 1.03 & 2.04 & 1.92 & 1.32 & 1.02 & 1.09 \\ 
\hline  
\end{tabular}
\end{table}

\section{Discussion} \label{sec.disc}

This paper proposes a robusified posterior for improving inference on a large number of parallel effects.  By providing significant protection against misspecified priors, our method encourages the use and specification of genuinely informative priors instead of defaulting to a weak and ineffective prior. For example, Method 6 in Section 4 can be an excellent choice if we believe that the tails of $\pi_0$ are between a short-tailed normal and a long-tailed t-distribution. Other approaches to enhance the robustness of Bayesian inference have been proposed in different contexts and models.  For example, \cite{Lazar:2003aa} replaces the likelihood function in the Bayesian posterior by an empirical likelihood, which achieves improved robustness by reduced specification in the likelihood.  Also, \cite{Hoff:2007aa} proposed to replace the likelihood of the complete data by the likelihood of the rank of the data to remove nuisance parameters in a semi-parametric copula estimation. The robustified posterior in this paper is specifically developed for estimating a large number of parallel effects.  By utilizing asymptotic behavior of order statistics and the unique structure of parallel effects, our method has the distinctive advantage of improving robustness with little or no loss of inferential efficiency even when the working prior is correctly specified.

Finally, we have previously proposed a rank-based robustified posterior in which the posterior of $\theta_i$ is computed conditioned on the rank of $y_i$ among $y_1,\ldots,y_p$ instead of the value of $y_i$ itself \citep{Liao:2014aa}. The rank-based posterior has similar properties as the robustified posterior in this paper but works well only when error $\varepsilon_i$ have similar variation across $i=1,\ldots,p$.  In contrast, the robustified posterior in this paper only requires the error distribution in \eqref{eq:canonical} to be continuous. 

\appendix 
\section{Proof of Theorem \ref{thm:sup}}
\begin{proof}
Formally, first consider the marginal distribution of order statistics $\tilde u$:
\begin{align*}
f(\tilde u \mid \pi_0) &= \int f(\tilde u \mid \by, \pi_0) f(\by \mid \pi_0) d\by.  
\end{align*}
where $f(\by\mid\pi_0) = \int f(\by\mid\theta) \pi_0(\theta) d\theta$ is the marginal distribution of $\by$. It follows from \eqref{eq:quantile} that $f(\tilde u \mid \pi_0)$ is the joint distribution of the order statistics from uniform $[0,1]$ \citep[see, e.g.,][p. 72]{Shao:1999aa}.
Let  $\tilde u$ be a draw from $
f(\tilde u \mid  \pi_0)$.  The Glivenko-Cantelli
theorem and the Berry-Esseen theorem state that, as $p \rightarrow
\infty$, the
empirical distribution of $\tilde u$ converges to the
function $F(x) = x$ uniformly on $x \in [0, 1)$.  Recent refinements to these theorems~\citep[Lemma 2]{Fresen:2011aa} are able to characterize the behavior of the
order statistics $u_{[i]}$ directly: \begin{equation} \label{eq.unif}
\sup_{1 \leq i \leq p} \left| u_{[i]} - \frac{i}{p+1}\right| \rightarrow
0 \end{equation} in probability.  

Now we show the asymptotics in \eqref{eq.unif}, derived under the marginal distribution $f(\tilde u \mid \pi_0)$, can be extended to the conditional distribution $f(\tilde u \mid y, \pi_0)$. It follows from equation \eqref{eq.unif} that, for any given $\delta_1>0$ and as $p\rightarrow\infty$, we have 
\[
\prob{\sup \left|u_{[i]}-\frac{i}{p+1}\right|>\delta_1}\rightarrow 0.
\]
Now for any $y$, define 
\[
D(y)\equiv \prob{\sup \left|u_{[i]}-\frac{i}{p+1}\right|>\delta_1 \mid y},
\]
where the right side is the conditional probability given $y$.  It follows that 
\[
\prob{\sup \left|u_{[i]}-\frac{i}{p+1}\right|>\delta_1}=E_y\left(D(y)\right),
\]
where the expectation on the right is with respect to $y\sim f(y\mid\pi_0)$.  Since $D(y)\ge0$ for every $y$ and $E_y(D(y))\rightarrow 0$, we have, for any $\delta_2>0$,
\[
\prob{D(y)>\delta_2}\rightarrow 0
\]
as $p\rightarrow \infty$, where the probability is evaluated with respect to $y\sim f(y\mid\pi_0)$.  In other words, except on a small set of $y$ whose probability goes to $0$, we have 
\[
\prob{\sup \left|u_{[i]}-\frac{i}{p+1}\right|>\delta_1 \mid y}\le \delta_2,
\]
for every $y$ when $p$ is sufficiently large.  
\end{proof}


\bibliographystyle{ECA_jasa}
\bibliography{robust}

\end{document}